\documentclass[12pt]{article}
\usepackage[T1]{fontenc}%
\usepackage[hmargin=1.5cm]{geometry}
\usepackage{graphicx}
\usepackage{color}

\begin{document}

\title{Scaling evidence of the homothetic nature of cities}

\author{R\'emi Lemoy and Geoffrey Caruso\\
%\small Universit\'{e} catholique de Louvain,
%   \small  \\
%   \small e-mail: {\tt hpotter@uclouvain.be}
%   \small University of Luxembourg, Maison des Sciences Humaines, \\
%   \small 11 Porte des Sciences, L-4366 Esch-Belval, Luxembourg\\
%   \small e-mail: {\tt remi.lemoy@uni.lu}
%\and Geoffrey Caruso\\
   \small University of Luxembourg, Maison des Sciences Humaines, \\
   \small 11 Porte des Sciences, L-4366 Esch-Belval, Luxembourg\\
%   \small Roundtable Institute\\
   \small e-mail: {\tt remilemoy@gmail.com, geoffrey.caruso@uni.lu}
   %\\ \small no e-mail
% followed by \and and other authors with affiliations if needed
}
\date{Draft working paper - version \today}

\maketitle
\vspace{-0.3cm}



\abstract{In this paper we analyse the profile of land use and population density with respect to the distance to the city centre for the European city. In addition to providing the radial population density and soil-sealing profiles for a large set of cities, we demonstrate a remarkable constancy of the profiles across city size.

Our analysis combines the GMES/Copernicus Urban Atlas 2006 land use database at 5m resolution for 300 European cities with more than 100.000 inhabitants and the Geostat population grid at 1km resolution. Population is allocated proportionally to surface and weighted by soil sealing and density classes of the Urban Atlas. We analyse the profile of each artificial land use and population with distance to the town hall.

In line with earlier literature, we confirm the strong monocentricity of the European city and the negative exponential curve for population density. Moreover, we find that land use curves, in particular the share of housing and roads, scale along the two horizontal dimensions with the square root of city population, while population curves scale in three dimensions with the cubic root of city population. In short, European cities of different sizes are homothetic in terms of land use and population density. While earlier literature documented the scaling of average densities (total surface and population) with city size, we document the scaling of the whole radial distance profile with city size, thus liaising intra-urban radial analysis and systems of cities. In addition to providing a new empirical view of the European city, our scaling offers a set of practical and coherent definitions of a city, independent of its population, from which we can re-question urban scaling laws and Zipf's law for cities.}

\section{Introduction}
This paper conducts a detailed empirical analysis of how the internal structure of the European city scales with its population, with the aim to start bridging the gap between two important strands of the urban modelling literature, namely the analysis of systems  of cities and the analysis of the internal structure of cities. We contend that space is insufficiently taken into account in the former, especially theoretically, while scaling effects are insufficiently considered in the latter literature, especially empirically. 

At the "systems of cities" (or urban scaling laws) end of the literature, there is simply no internal spatial structure. The research is mainly empirical and focuses on how different aggregate variables that describe the socio-economic or physical state of cities evolve with city population (recent critical examples are \cite{bettencourt2007, shalizi2011, louf2014, leitao2016}). Assuming only a single aggregate attribute and an average density for each city is an hazardous assumption. Indeed, and fortunately, a large portion of urban economics and urban geography research is devoted to the variation of densities, land uses, income, or rents within cities because they matter for understanding and planning cities. Especially important to this internal heterogeneity are distance to the center effects, i.e. radial effects, including transport costs but also centre-periphery amenities, which impact land use and population densities, as shown along the Alonso-Muth-Mills monocentric tradition \cite{alonso, muth, mills, Fujita89}.

Notwithstanding the relevance of internal city structures, the theory (which we denote hereafter "systems of cities" or "urban systems") capable of approximating the distribution of urban populations (particularly Zipf's law) and other city aggregates across cities is largely dominated by an \textit{a-spatial} random growth framework (following Gibrat, see for example \cite{Krugman96, GabaixIoannides04, FavaroPumain09GA}) that sometimes invokes endogenous intra-urban - but usually non-explicitly spatial - processes. For example, \cite{Gabaix99AER} and \cite{Eeckhout04AER} relate positive agglomeration and negative externalities effects to total population, not to the internal distribution of this population. \cite{rozenfeld09NBER} endogenise land consumption in a random growth model but assume costs to be proportional to the total surface consumed and densities to be constant. \cite{bettencourt2013} proposes an intra-urban dissipative model with social interactions and transport networks, and relates outputs to scaling laws, but ignores radial distance effects. %The interactions suggested by Bettencourt however are not economically explicit nor empirically validated. %The model is inspired by physical growth processes that echoes earlier research on the internal structure of cities using fractals (e.g Frankhauser94 and Longley Battyxx, see Batty comment SCIENCE).

%Gabaix, Xavier. 1999. "Zipf's Law and the Growth of Cities." American Economic Review, 89(2): 129-132. 
 %Eeckout, Gibrat's Law for (All) Cities. The American Economic Review. 94, 5, 1429-1451

At the intra-urban scale end of the literature, the empirical validation of increasingly refined theoretical models from the monocentric tradition \cite{alonso, muth, mills} is still very rare. Most of the literature develops from stylised facts without validating, jointly, all model components (density and rent profiles, costs, income, urban fringe). \cite{Ahlfeld11JRS} offers one of the most comprehensive tests for a single city (Berlin). Only the analysis of population density gradients, since \cite{clark1951}, received an empirical investigation effort that is comparable to the urban scaling literature, though more fragmented (the latest exhaustive review to our knowledge is \cite{McDonald89}). It usually considers individual case studies, not cross-sections of numerous cities. %Glaeser and Kahn 131 cities ? Check what they do.
 More importantly, population density profiles are rarely discussed against land use profiles, the limits of the city, and total urban population. This is particularly sad because the location of the urban fringe (the maximum extent of urban uses) and the total population are key outcomes (or constraints) in monocentric models. In Europe \cite{guerois2008} fitted a double linear function to the profile of urbanised surfaces for 40 cities but could not link it to density profiles due to the absence of coherent population data and definition of cities. In the US, \cite{mcgrath05JUE} related the urbanisation area (via the urban fringe) to essential parameters of the standard monocentric model (income, transport costs and the opportunity cost of land) as well as to total population for 30 cities, but disregarded the density profile.

Obviously data gathering is a strong limiting factor to a thorough validation of standard monocentric models. This is however gradually changing with the emergence of more precise, comparable and almost ubiquitous land use data.

An important difficulty that remains in both strands of the literature for comparing results is the definition of spatial units. Within cities, density measurement are affected by MAUP effects as soon as areal units (census tracts, municipalities) are used to retrieve population or artificial land surfaces. Hence the importance of using the finest possible geographical units and both land use and population data. In addition, the chosen limit of the city itself influences the value of aggregate attributes and resulting scaling laws (e.g. \cite{louf2014, arcaute2015}). Some researchers use administrative units and a functional definition of cities (e.g. based on commuting patterns), other consider raster data and morphological criteria (continuity of built-up space) or population density cutoffs. There is need for comparable %and possibly endogenous
 methods for defining cities.

\cite{Rozenfeld2008PNAS, rozenfeld2011, arcaute2015}  define the boundaries and surface of cities using a clustering algorithm in order to analyse deviations to Gibrat's law. Interestingly, \cite{rozenfeld2011}, using this endogenous definition of cities, find that areas are essentially proportional to population with almost no role left to density within the scaling behaviour. In the meantime the authors also "defer to later work the interesting question of the heterogeneity [in density] within cities"(\cite{rozenfeld2011}, p.2221). We therefore find ourselves with a theory of urban scaling that is at best devoted to small density variation across cities - a variation acknowledged to be small %(see also meta analysis of Melo et al)
 empirically compared to intra urban variations,  while an entire empirical and theoretical literature is devoted to this variation but struggles to assemble pieces empirically, particularly across the city size distribution. We are somehow only left with the intuition of Nordbeck (45 years ago, \cite{nordbeck1971}) that cities are simple homotheties and have the same form and shape whatever their size.

Our paper challenges this situation by providing, for Europe, an analysis of the scaling of the entire population density and land use profiles. Our work is grounded on the finest comparable land use and population data. Although our approach relies on an \textit{ex-ante} definition of cities (European Larger Urban Zones), the scaling relationship we find - remarkably rejoining Nordbeck's ratios - allows us to derive a new definition of cities that is consistent across sizes. In addition to strongly backing-up the monocentric nature of the European city and a very robust population density profile, we can therefore also verify aggregate scaling behaviour (Zipf) for different definitions of cities.

%Let us note in this line of research the pioneering work and powerful intuitions of Nordbeck \cite{nordbeck1971}, who used a closely related approach and detailed Swedish data from the 1960s to study the scaling relationship between the area and population of a city. He obtained closely related and coherent results, as discussed further.

%Our main objective is to retrieve generic laws that can support the calibration of monocentric urban economic models for European cities. In this approach, the city is observed in a very coarse-grained way, with the objective of comparing reality to generic models, irrespective of local specificities. We study only the evolution of variables (like land use or population density) with respect to the distance to the city centre, considering for simplicity that these variables present a symmetry of revolution around the centre. We call this approach monocentric as a reference to the standard urban economic model, but we could also name it radial, as the distance to the centre (radius) is the only variable determining the location of studied areas.
%We study the evolution with distance from the city center of the quantity of land used for different purposes: housing, roads, railways, urban green, water, agriculture, forest.

\section{Data and methods}

This work uses as a main data source the European Urban Atlas 2006 developed by the GMES/Copernicus land monitoring services. The database provides a precise description of land use at 5m resolution in the 300 major European urban areas (defined according to Eurostat Larger Urban Zones (LUZ)) with more than 100.000 inhabitants in 2006. These urban areas make up more than 200 million inhabitants, i.e. more than 40\% of the EU population in 2006. As in any system of cities, many of these 300 urban areas have a small population, and few have a high population. This fact has been linked to Zipf's law for many systems (see the review by \cite{nitsch2005}) and has been studied for European cities by many authors including recently \cite{bettencourt2016}.

\begin{figure}[h!]
\begin{center}
\begin{tabular}{cc}
\includegraphics[height=6.0cm]{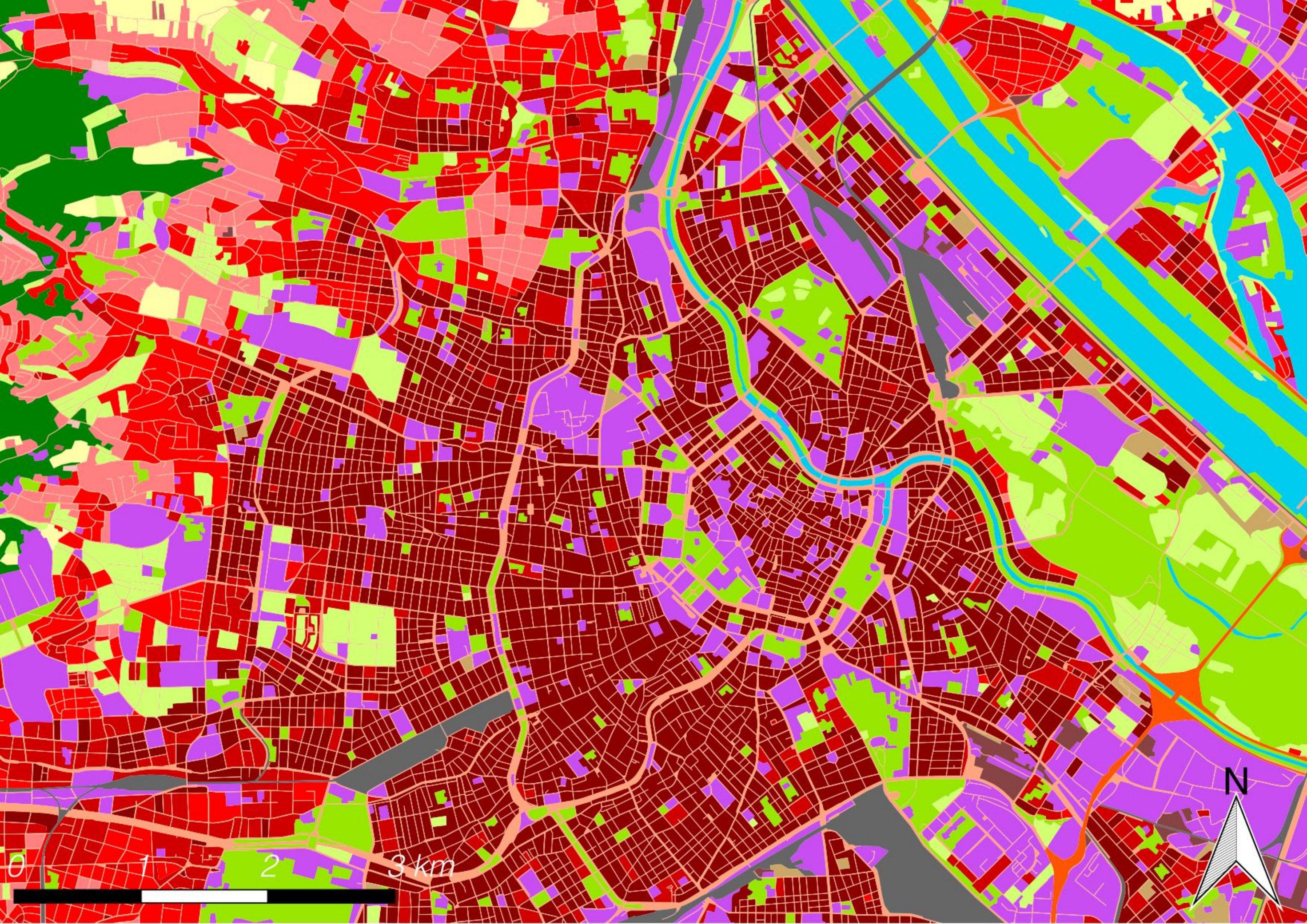} &
\includegraphics[height=6.0cm]{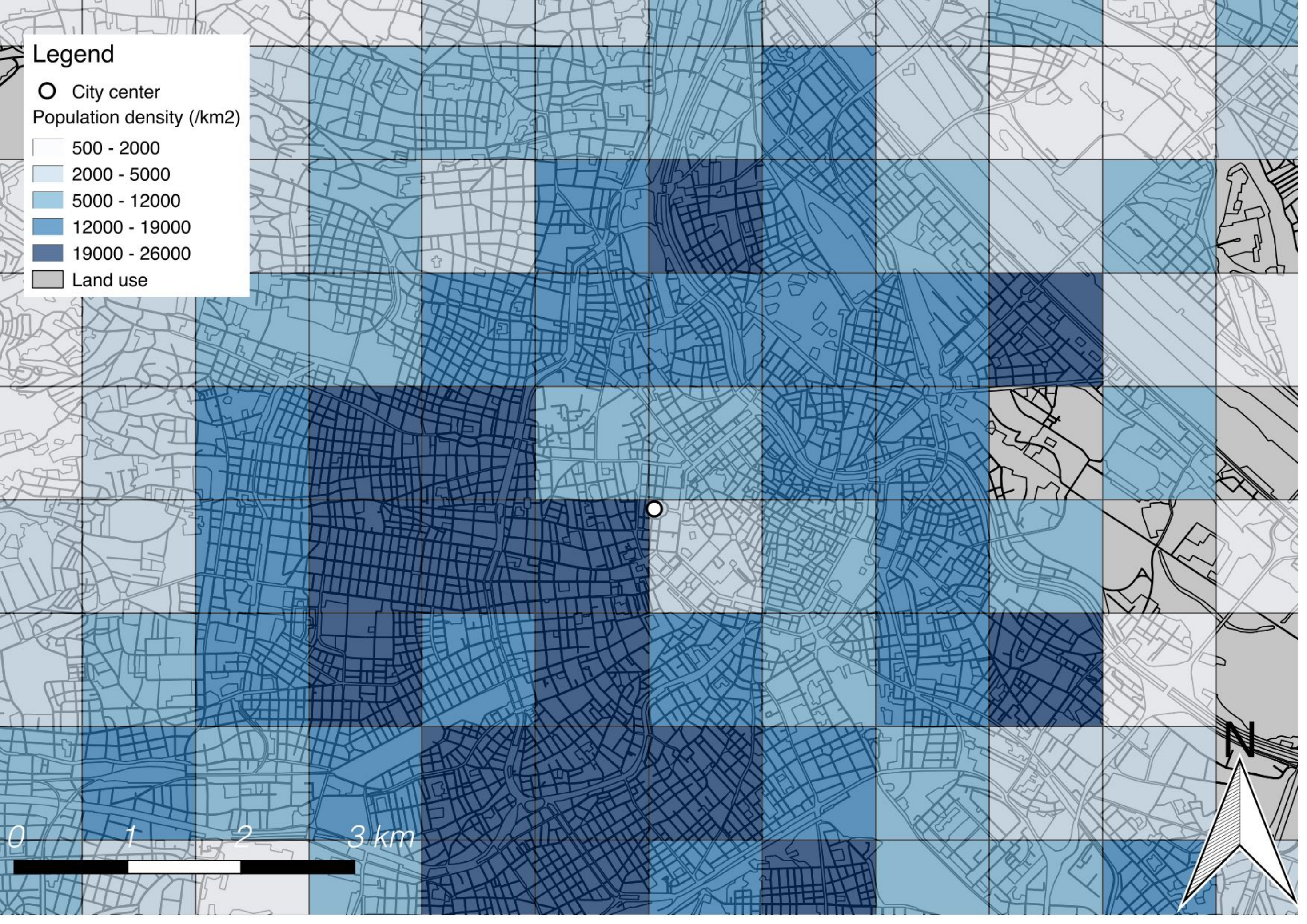} \\
\includegraphics[height=6.0cm]{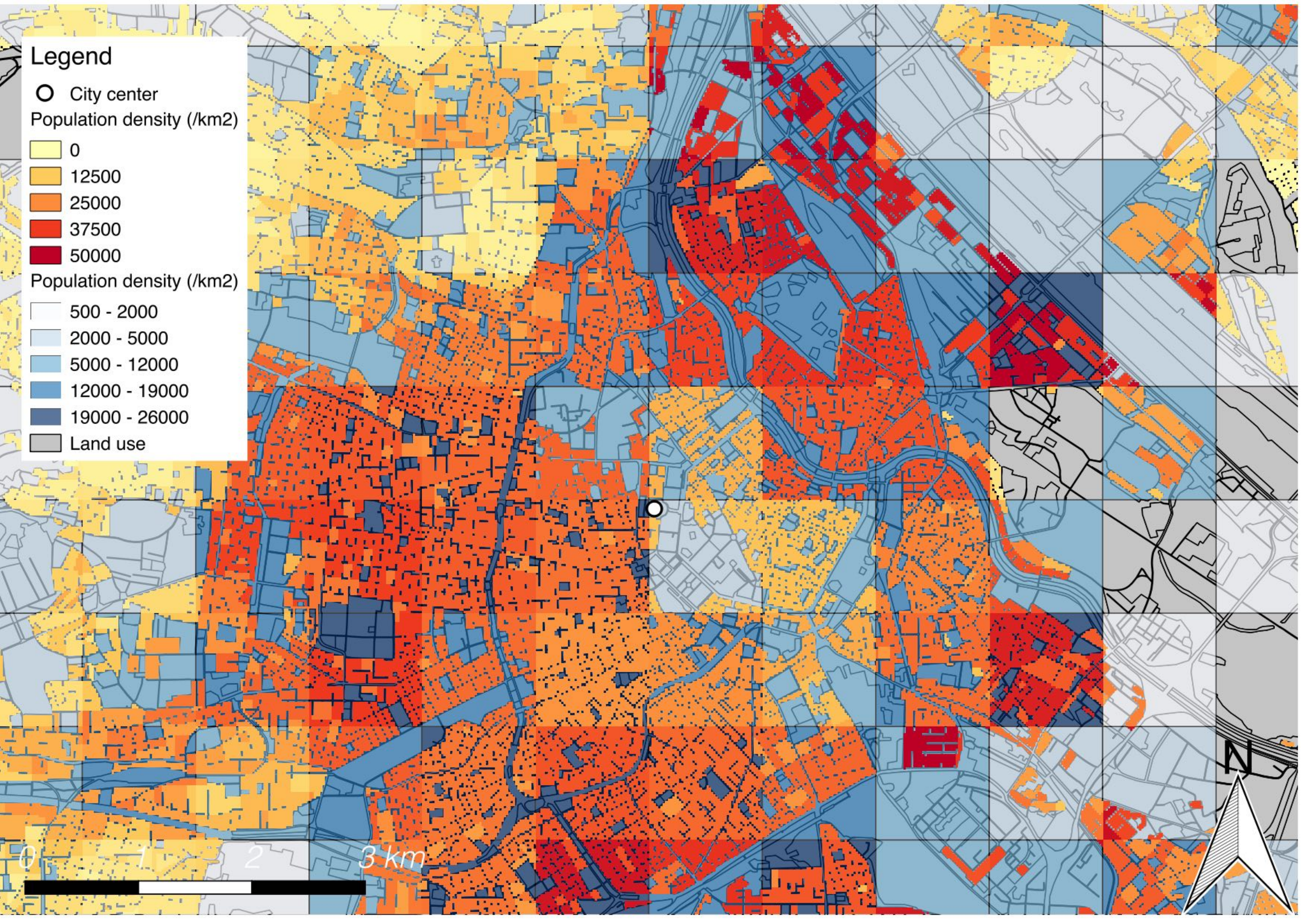} &
\includegraphics[height=6.0cm]{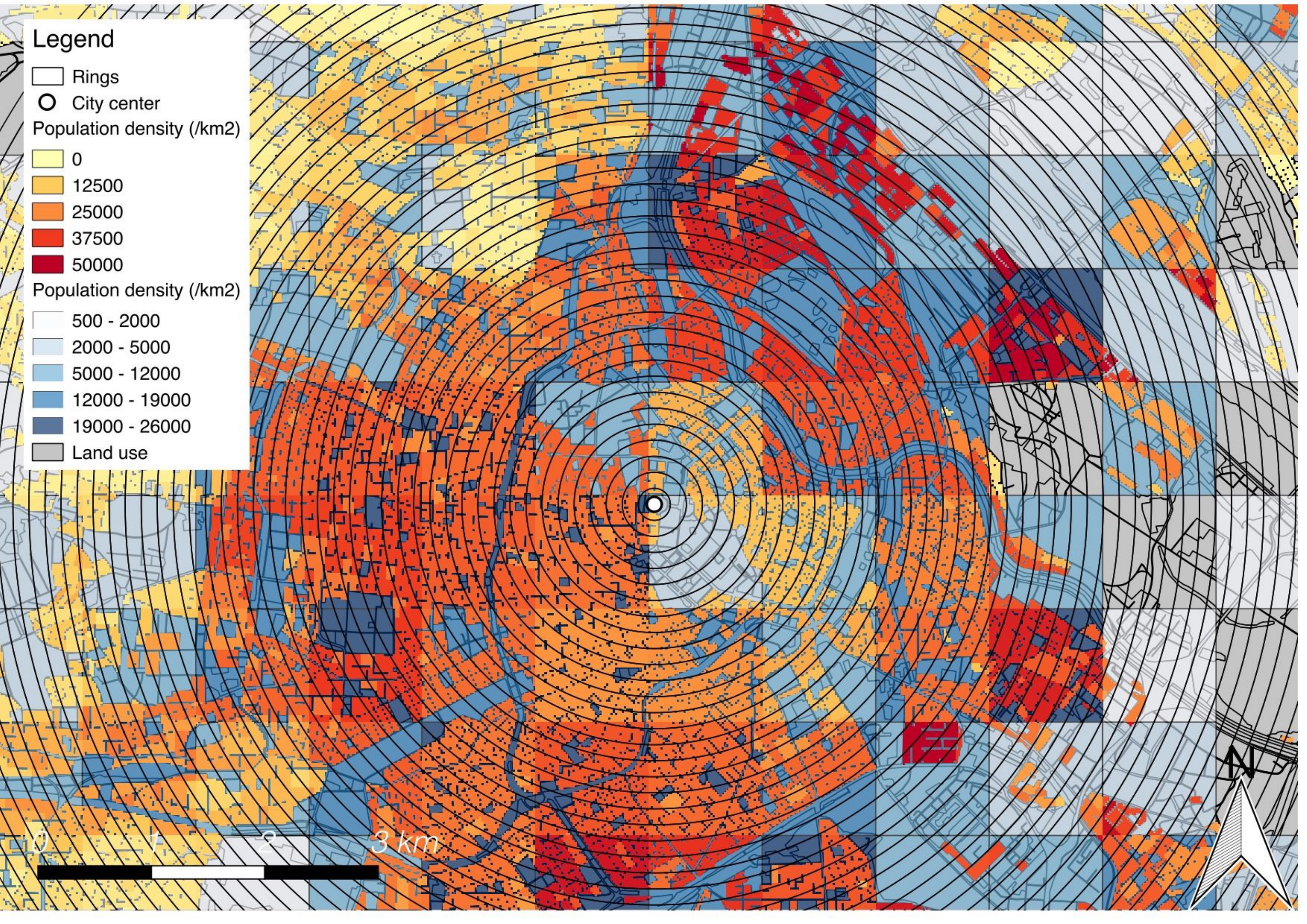} 
\end{tabular}
\caption{{\small Illustration of the datasets and methods on Vienna, Austria: Urban Atlas 2006 (top left), Geostat population grid (top right). Distribution of the population according to land use (bottom left). Rings of the monocentric analysis (bottom right).}}
\label{data}
\end{center}
\end{figure}

We transform the Urban Atlas dataset into a 20m resolution grid with the same land use information using a central point rule for aggregation. This transformation facilitates computing while preserving spatial objects of small width, like main roads (conversely to a majority rule). We then combine this land use grid dataset with population density from the Geostat population grid, which covers the European Union (EU) with a 1km resolution, also for the year 2006. Population counts given by the Geostat grid are downscaled to the 20m land use grid, using the "urban fabric" classes of the Urban Atlas. Those classes are based on different levels of soil sealing (S.L.) as follows and illustrated on Figure \ref{data} for the inner city of Vienna, Austria:
%\begin{itemize}
%\item Continuous Urban Fabric (S.L.: > 80\%)
%\item Discontinuous Dense Urban Fabric (S.L.: 50\% - 80\%) 
%\item Discontinuous Medium Density Urban Fabric (S.L.: 30\% - 50\%) 
%\item Discontinuous Low Density Urban Fabric (S.L.: 10\% - 30\%) 
%\item Discontinuous Very Low Density Urban Fabric (S.L. < 10\%) 
%%\item Isolated Structures
%\end{itemize}
Continuous Urban Fabric (S.L.: > 80\%), Discontinuous Dense Urban Fabric (S.L.: 50\% - 80\%), Discontinuous Medium Density Urban Fabric (S.L.: 30\% - 50\%), Discontinuous Low Density Urban Fabric (S.L.: 10\% - 30\%), Discontinuous Very Low Density Urban Fabric (S.L. < 10\%). 
%\item Isolated Structures
Each class was given a weight, respectively 0.85, 0.65, 0.4, 0.2 and 0.05, describing its contribution to the residential location of households. The population in each 1km cell is then attributed to the urban cells using these weights.

Our aim is to conduct a radial analysis of the land use and population density profiles and study how theses curves scale with respect to total city population. An important choice therefore is the location of the city centre. In this work, we choose the location of the city hall, which corresponds well in most European cities to the historical center of the city, and is usually located in a completely artificialised close to high population densities. Other authors have use the coordinates of the city hall as the location of the city centre (see e.g. \cite{wilson2012, walker2016}). Appendix \ref{app1} shows that the precise location of the city center has only a small influence on the results.

\begin{figure}[h!]
\begin{center}
\includegraphics[height=9.0cm]{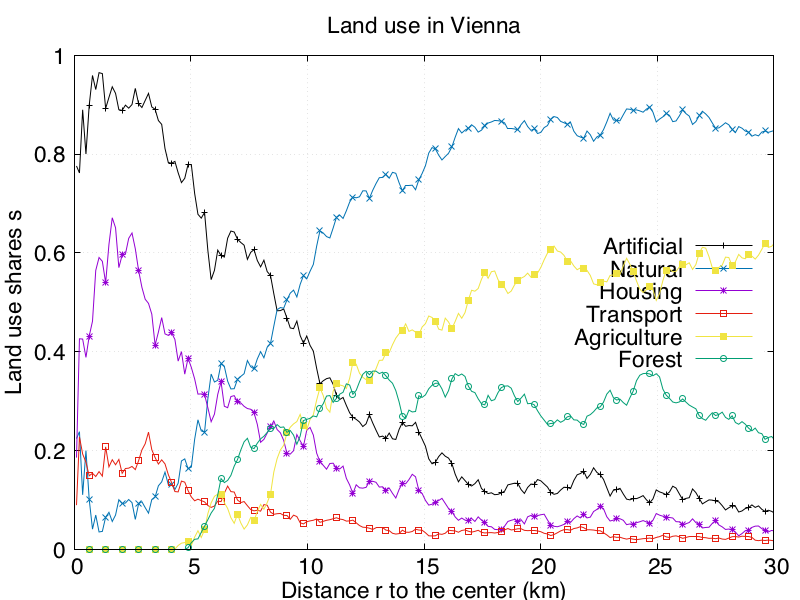}
\caption{{\small Land use shares as functions of the distance to the center in Vienna.}}
\label{vienna}
\end{center}
\end{figure}

We then define concentric rings of fixed width $100\sqrt{2}\simeq 141$ m around the city hall, and average population density and the share of each land use within each of these rings (figure \ref{data}).
Looking a these curves individually shows that cities are very comparable, with a concentration of artificial land uses (mostly urban fabric and transport infrastructures) around the center, and increasingly natural land uses (agricultural or forest) as one moves away from the city center. This is illustrated on Figure \ref{vienna} for Vienna, Austria.

\section{Scaling laws}

In order to compare cities of different size and to identify whether there exists a general radial profile of urbanised land use and density that would be representative of a European city, we study the scaling of the artificial land use share and population density curves.

As we recall in our introduction, the city population has been widely used as a scaling parameter, to evaluate the evolution with city size of different variables, such as income or road space. In line with this literature, we use the total population $P_c$ of the Larger Urban Zone (LUZ) as a scaling parameter and a measure of the size of the urban area $c$. %\textcolor{red}{JE NE VOIS PAS TROP LE BUT DE LA PHRASE SUIVANTE}\textcolor{blue}{Another reason for this choice is that a city, as illustrated by the AMM monocentric model, is fundamentally characterised by a concentration of residential locations for humans, which the population size measures.}
 For consistency we compute total population for each urban area from the population grid data (sum over all cells of the LUZ).

\subsection{Artificial land} 

\begin{figure}[h!]
\begin{center}
\begin{tabular}{cc}
\includegraphics[height=6.7cm]{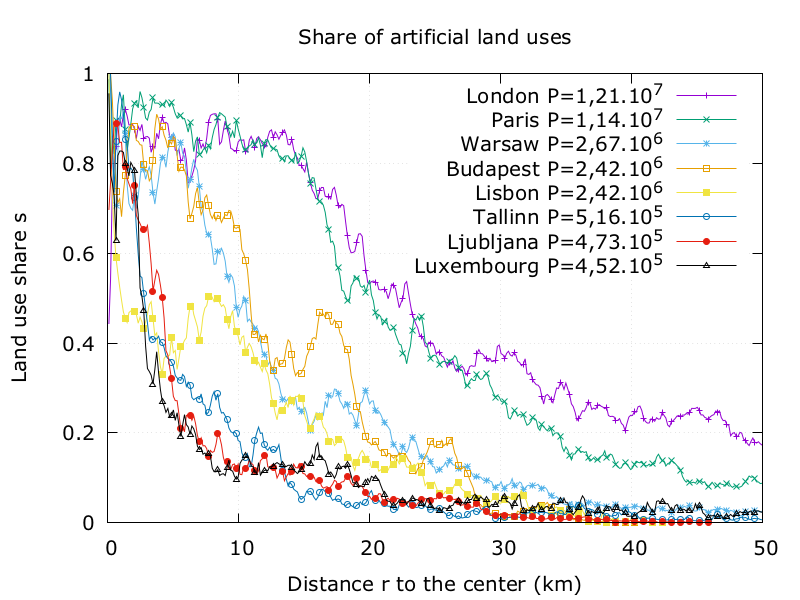} &
\includegraphics[height=6.7cm]{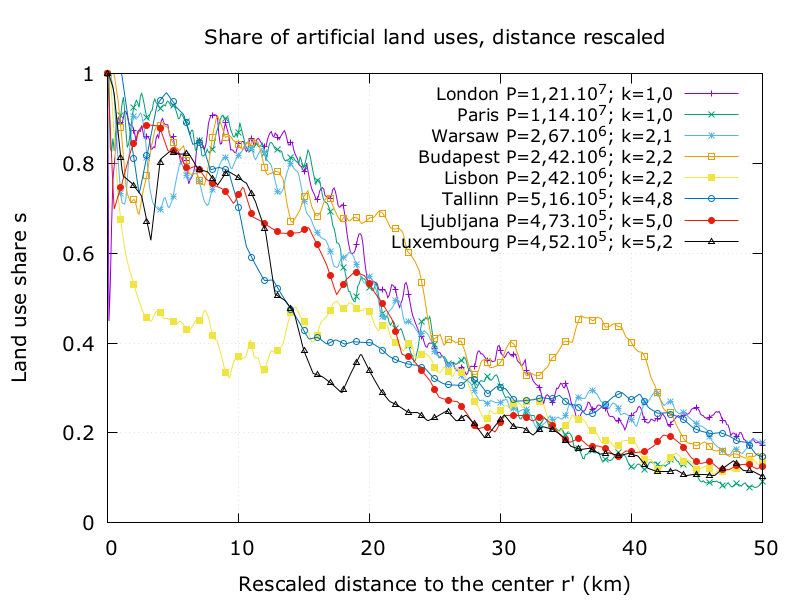}
\end{tabular}
\caption{{\small Left: shares of artificial land use as functions of the distance to the center in different European capital cities $c$. The population $P_c$ is given in the legend. Right: rescaled curves for the same cities. The rescaling factor $k_c$ is given in the legend.}}
\label{artif}
\end{center}
\end{figure}

We denote by $s^c_a(r)$ the share of artificial land use\footnote{By "artificial" we mean all land use categories of the Urban Atlas except water, agriculture, urban green and forests.} at distance $r$ of the city center in city $c$. When comparing the shape of  $s^c_a(r)$ curves for different cities $c$, as shown on the left panel of figure \ref{artif} for different European capital cities, one can observe that their evolution is similar: they start at roughly 100\% in the very center and decrease further from the center. Additionally, we see that this decrease is faster when the  the city population is smaller. On the left panel of figure \ref{artif}, the different curves appear clustered around population size groups: the two biggest cities in the dataset (London and Paris, with roughly 12 million inhabitants), three smaller capital cities (Warsaw, Budapest, Lisbon, roughly 2.5 million inhabitants) and three small ones (Tallinn, Ljubljana and Luxembourg, with roughly 500 000 inhabitants). In order to compare those curves to each other, it is then interesting to rescale the horizontal axis proportionally to a power of the city population $P_c$.
We perform this on the right panel of figure \ref{artif}, using on the horizontal axis the rescaled distance $r'=r/\sqrt{P_c/P_{London}}$. The legend of the figure indicates the population size $P_c$ and the rescaling parameter $k_c=\sqrt{P_c/P_{London}}$ for each city $c$. We use London, the largest city of the dataset, as a reference: $k_{London}=1$ (and also $k_{Paris}\simeq 1$ because Paris is nearly as big as London). Figure \ref{artif} shows that with this rescaling, the different curves of artificial land use share $s^c_a(r)$ look very much like each other, even for cities of very different size. We note that Lisbon for example deviates from that rule close to the city centre, on both panels of figure \ref{artif}. This can be explained by the fact that Lisbon is a coastal city (on the Atlantic Ocean) and also lies on the Tagus River estuary, which occupies a large space very close to the city centre.

\begin{figure}[h!]
\begin{center}
\begin{tabular}{cc}
\includegraphics[height=7cm]{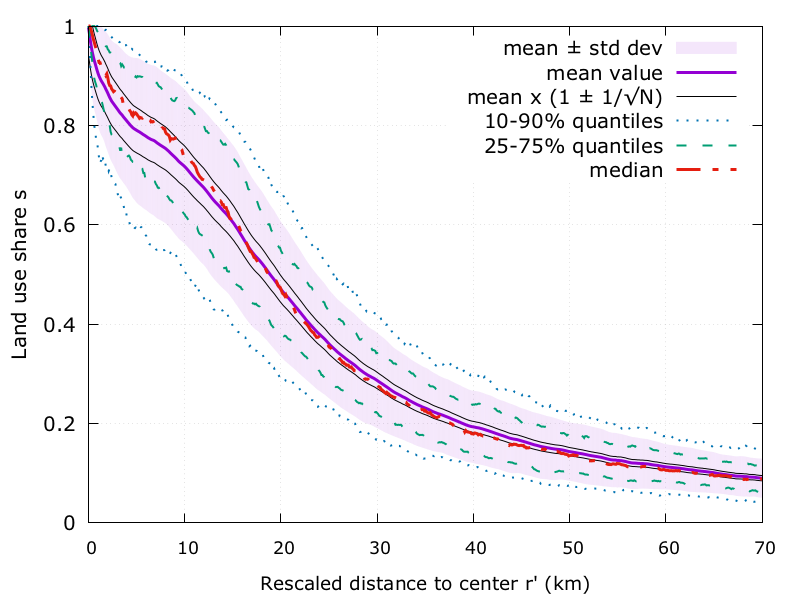} &
\includegraphics[height=7cm]{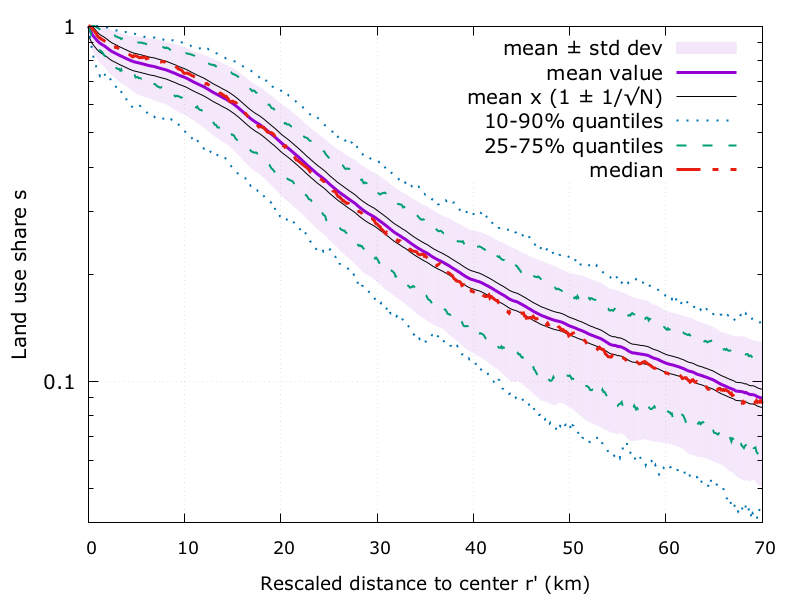}
\end{tabular}
\caption{{\small Average artificial land use share $s_a(r')$ and fluctuations around the average, as functions of the rescaled distance to the center $r'$, with linear axes (left) and logarithmic y-axis (right).}}
\label{resc-artif}
\end{center}
\end{figure}

\begin{figure}[h!]
\begin{center}
\includegraphics[height=9.0cm]{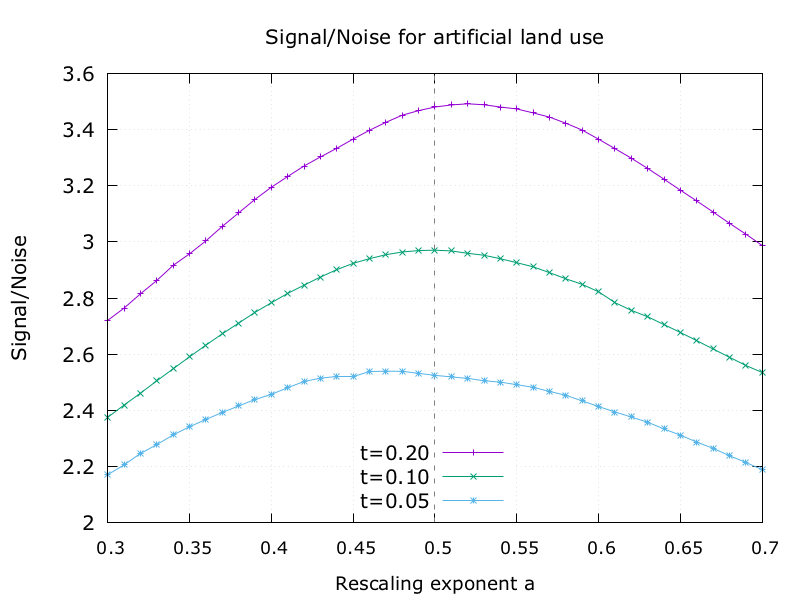}
\caption{{\small Signal over noise ratio for the artificial land use share, with different values of the rescaling exponent $a$ and different values of the integration threshold $t$. The dashed vertical line indicates $a=1/2$.}}
\label{exp_artif}
\end{center}
\end{figure}

Figure \ref{resc-artif} then presents the average curve for all $N=300$ cities of the database, as well as fluctuations around the average. Compared with $1/\sqrt{N}$ relative fluctuations, the fluctuations observed are quite high, as can be expected from cities with very different characteristics in addition to sizes (geography, climate, history, economy, planning policy, culture etc.) but the rescaling still captures a very clear common trend. In order to obtain data at common rescaled distances $r'$ despite the rescaling for figure \ref{resc-artif}, we use a linear interpolation between the closest data points. The average, standard deviation and quantiles are computed over all cities at each rescaled distance $r'$.
One can wonder if the rescaling used here, with the square root of city population, is the optimal one in order to obtain comparable curves for cities of different sizes. We propose next to compare different rescaling exponents $a$, used in the distance rescaling $r'=r/(P_c/P_{London})^a$. We measure the performance of the rescaling for each exponent using a signal over noise ratio: the average artificial land use $\langle s_a(r)\rangle$ is the signal, and the standard deviation $\sigma(s_a(r))$ of the artificial land use is the noise. We compute the average value of this signal over noise ratio between $r'=0$ and $r'$ such that $\langle s_a(r)\rangle=t$, where $t$ is an artificial land use share threshold. Figure \ref{exp_artif} shows the evolution of this average signal over noise ratio with the rescaling exponent $a$, for different values of the threshold parameter $t$.
One can observe that the exponent $a$ yielding the highest signal over noise ratio is extremely close to $1/2$, which is the exponent we used so far, for different values of the threshold parameter $t$.

How can we interpret this rescaling with the square root of city population? We have to remember that the land use curves represent averages over two dimensional rings, so that two horizontal dimensions are actually rescaled with respect to the square root of city population. This means that the artificial surface around the city center scales linearly with city population: cities are pretty much homothetic in terms of land use, and the scaling with population size is linear. This is also an empirical evidence of the monocentricity and the similarity of European cities, irrespective of their size or other characteristics. This result can be related to the proportionality between area and population found by \cite{rozenfeld2011} for US and UK cities using a clustering algorithm. It is also in line with the much earlier analysis of \cite{tobler1969}.

\begin{figure}[h!]
\begin{center}
\begin{tabular}{cc}
\includegraphics[height=7cm]{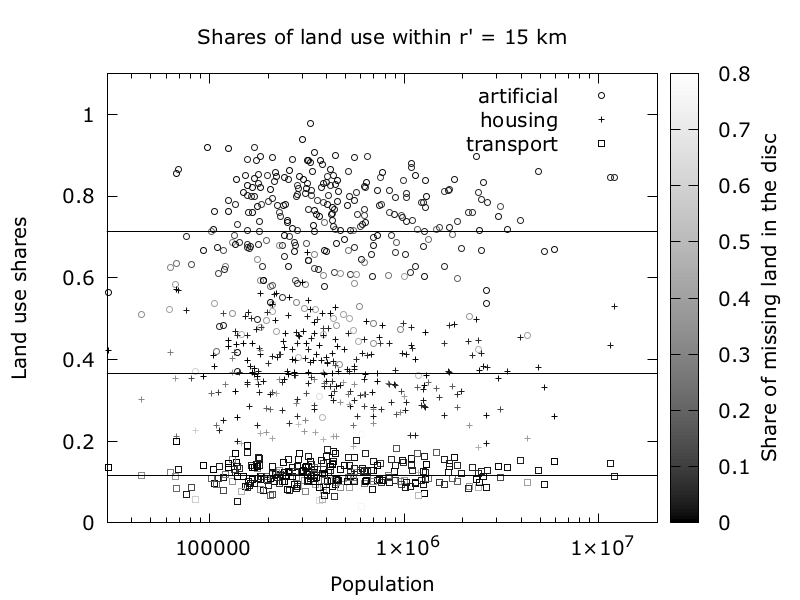} &
\includegraphics[height=7cm]{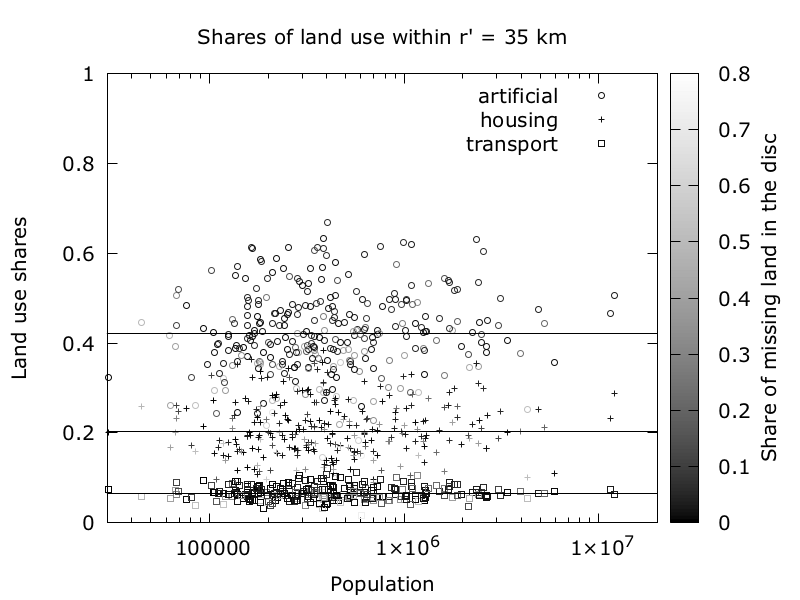}
\end{tabular}
\caption{{\small Average land use shares in discs of rescaled radius $r'=15$ km (left panel) and $r'=35$ km (right panel), for all cities of the database. The lines representing average land use shares are meant as guides to the eye.}}
\label{disc_artif}
\end{center}
\end{figure}

Figure \ref{disc_artif} gives a different perspective on the same phenomenon. From the previous results, we concluded that most European cities of the database are comparable in terms of land use as long as the distance to the center is rescaled with respect to the square root of population size. Choosing a specific distance to the center, for instance $r'=15$ km or $r'=35$ km, this rescaling defines discs of different sizes for each city, in which land use should be comparable.
Figure \ref{disc_artif} shows that the land use shares in these rescaled discs are indeed quite constant with respect to city population. Moreover, not only the artificial land use share is constant. The share of land attributed to "urban fabric" in the database, which corresponds mostly to housing, is also constant. And so is the share of land used for transport, which combines space used for roads and railroads. Fluctuations between cities are important, but the data does not present any drift from constant land use shares. Additionally, figure \ref{disc_artif} gives also a reason why some cities do not follow the general rule. Indeed, for many cities some land is missing in the studied discs of fixed rescaled radius $r'$: in most cases this missing land is covered by water, as many cities of the database are coastal cities. In some rarer cases we have border effects: land is missing because it is outside of the Larger Urban Zone of the considered urban area, either because it corresponds to another country (city on a border) or another urban area (neighbouring city).

\subsection{Population density}

\begin{figure}[h!]
\begin{center}
\begin{tabular}{cc}
\includegraphics[height=6.7cm]{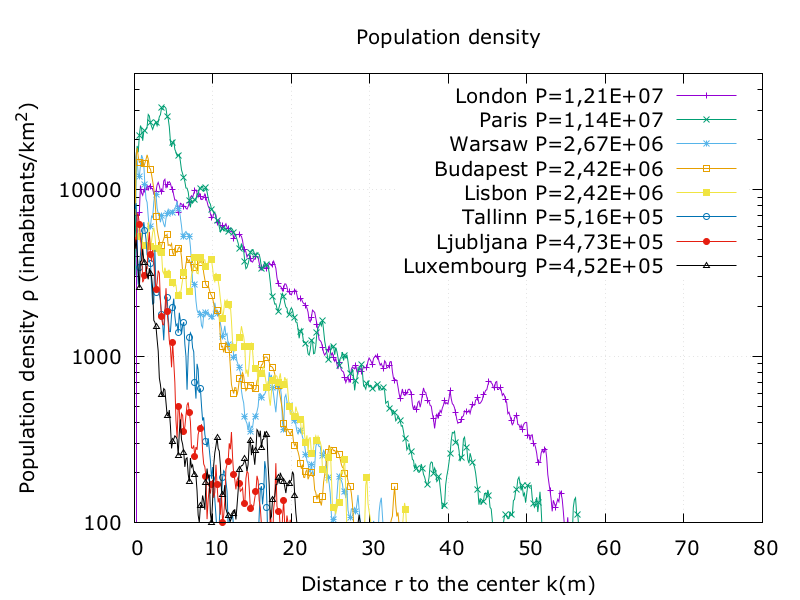} &
\includegraphics[height=6.7cm]{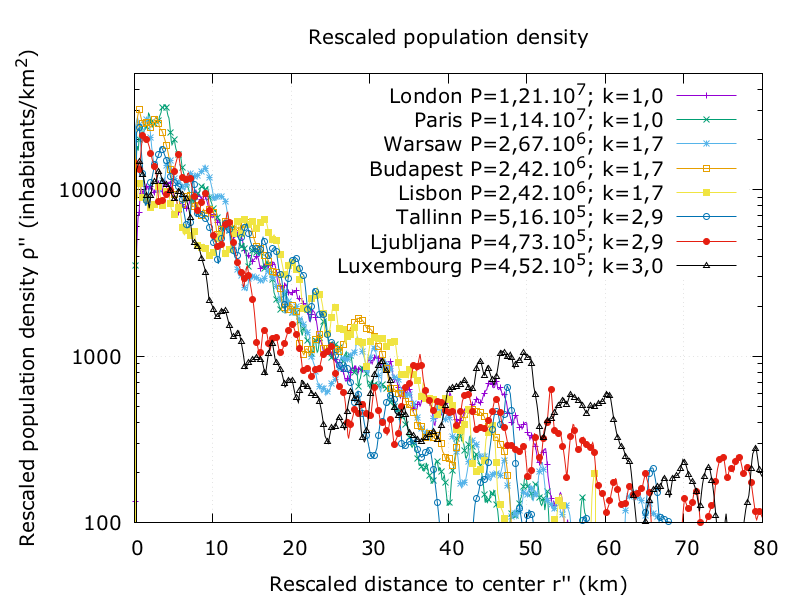}\\
\includegraphics[height=6.7cm]{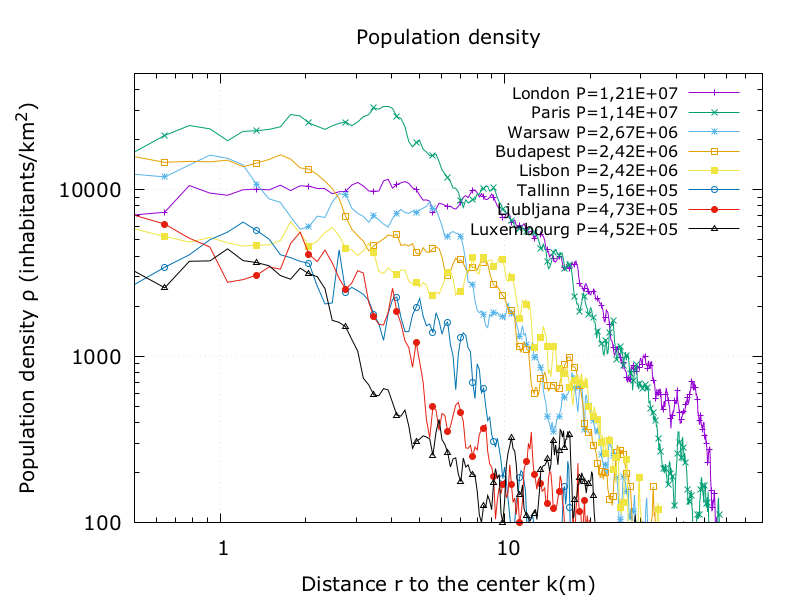} &
\includegraphics[height=6.7cm]{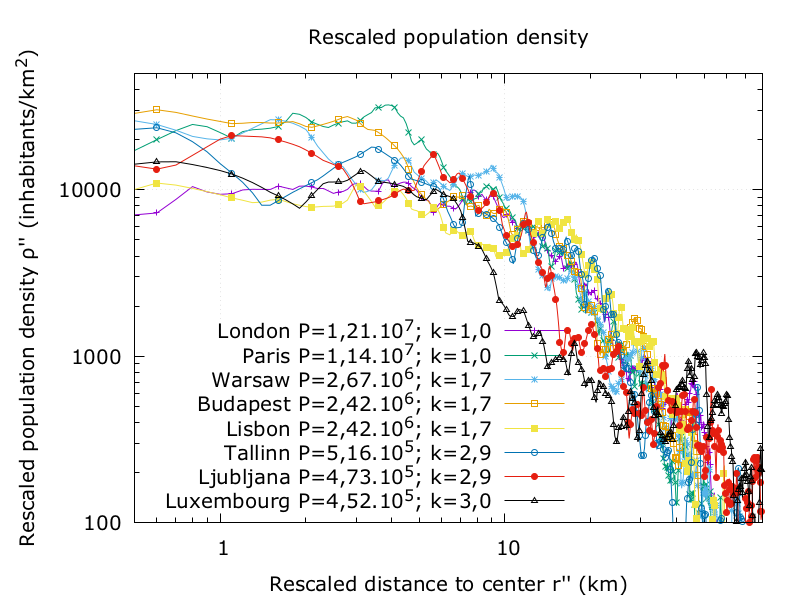}
\end{tabular}
\caption{{\small Left: population density as a function of the distance to the center in different European capital cities $c$. The population $P_c$ is given in the legend. Right: rescaled curves for the same cities. The rescaling factor $k_c$ is given in the legend.}}
\label{dens}
\end{center}
\end{figure}

We now perform a similar analyis for population density. The left side of figure \ref{dens} shows that the density $\rho(r)$ as a function of the distance $r$ to the center has a similar behaviour for cities of different sizes. It is a decreasing curve, very roughly exponential, except in the very center where it is relatively constant\footnote{Let us note that the original 1km resolution data can bias the results for very small distances to the center.} or even increasing (this is known since \cite{newling1969}). The decrease is faster for smaller cities, and the density in the center is higher for bigger cities. This suggests to rescale both horizontal and vertical axes together with respect to the city population.

We find that the exponent that works well in this case is the cube root of city population: we use a rescaled distance to the center of city $c$ noted $r''=r/(P_c/P_{London})^{1/3}$, and a rescaled density $\rho''(r'')=\rho(r'')/(P_c/P_{London})^{1/3}$.
The right side of figure \ref{dens} shows that with this rescaling, the density curves of cities of different sizes are quite similar, irrespective of their different characteristics. The figure indicates the rescaling factor $k_c$, which this time is $k_c=(P_c/P_{London})^{1/3}$.

\begin{figure}[h!]
\begin{center}
\begin{tabular}{cc}
\includegraphics[height=7cm]{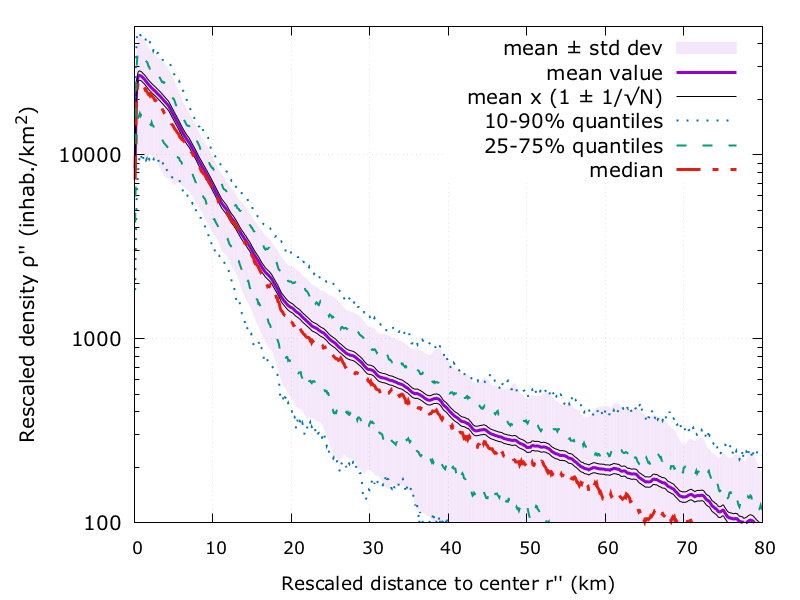} &
\includegraphics[height=7cm]{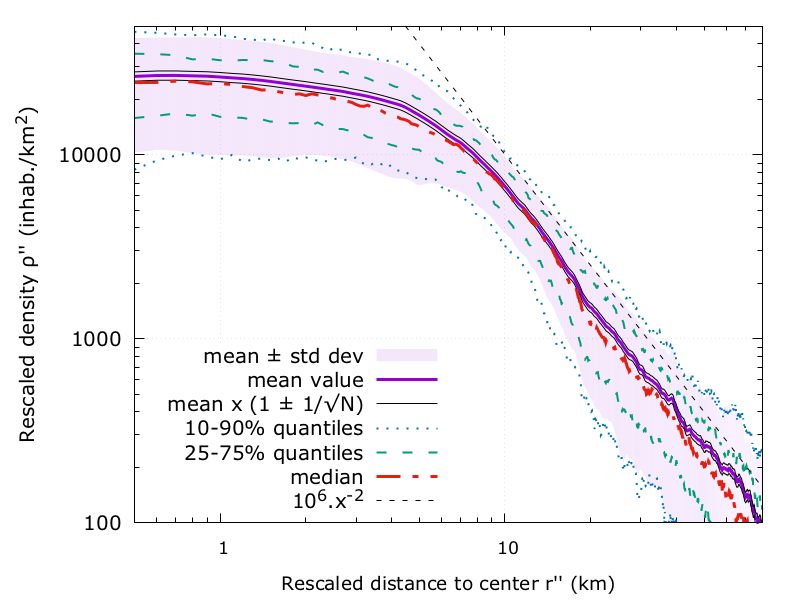}
\end{tabular}
\caption{{\small Distribution of the rescaled density $\rho''$ as a function of the rescaled distance to the center $r''$ with logarithmic y-axis (left) and log-log axes (right).}}
\label{resc-dens}
\end{center}
\end{figure}

Furthermore, figure \ref{resc-dens} presents the average (over all 300 cities of the database) rescaled density $\rho''$ at each rescaled distance to the centre $r''$, and the fluctuations around this average.
This figure shows that the negative exponential function, which has been widely used in the literature to describe the density curve of different cities since \cite{clark1951}, could fit only roughly this average density curve. Its shape could be more precisely described using for instance (the minimum of) two functions: one power-law with an exponent close to $-2$ would describe the density curve in the outskirts of the urban area, starting at $r''=10$ or 15 km, and a negative exponential (or another power-law with an exponent much closer to 0) for the central part of the city.

In a separate analysis not presented here, we check that the curves of figure \ref{resc-dens} remain roughly unchanged when different weights are used for averaging and quantiles. For instance, giving each city a weight corresponding to its population (which is what \cite{leitao2016} call "Person model") does not yield notable change for the curves of figures \ref{resc-artif} and \ref{resc-dens}, but we note that the compatibility of the density curve with a negative exponential is slightly improved.

One can here again wonder if the cube root of the city population is the right rescaling exponent for the density curve. The same signal over noise ratio as previously can be computed here, for different values of the rescaling exponent $b$ of both the rescaled distance to the center $r''=r/(P_c/P_{London})^{b}$ and the rescaled density $\rho''(r'')=\rho(r'')/(P_c/P_{London})^{b}$. The signal over noise ratio is averaged from $r''=0$ to a rescaled distance $r''$ such that the rescaled density reaches a threshold $t$ (expressed in inhabitants/km$^2$), $\rho''(r'')=t$.
\begin{figure}[h!]
\begin{center}
\includegraphics[height=9.0cm]{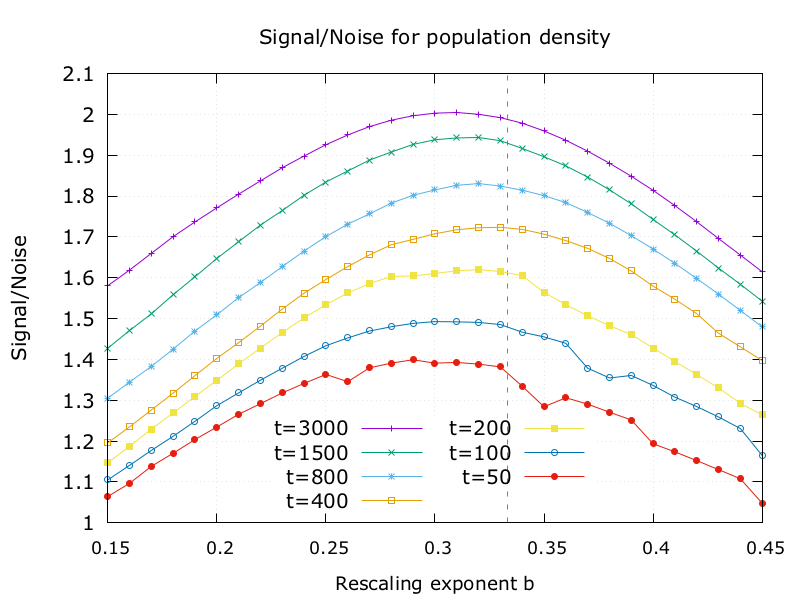}
\caption{{\small Signal over noise ratio for population density, with different values of the rescaling exponent $b$ and different values of the integration threshold $t$ (in inhabitants/km$^2$). The dashed vertical line indicates $b=1/3$.}}
\label{exp_dens}
\end{center}
\end{figure}
Figure \ref{exp_dens} shows that the exponent yielding the highest signal over noise ratio is close to $1/3$, for different values of the density threshold $t$. This result is again an empirical evidence of the monocentric or radial organisation of European cities, this time in terms of population density, and of their similarity across different sizes. It also shows that the population is located in 3 dimensions, and that bigger cities are also taller ones. An interesting perspective of work consists in applying this same approach to 3-dimensional data featuring building height (increasingly available for scientific purposes), in order to see how the volume of city buildings scales with population size. We can expect this scaling to be 3-dimensional, consistently with the scaling of population density studied here. We note also that this result is very consistent with early findings of Nordbeck \cite{nordbeck1971}, for whom "It seems legitimate to claim that all urban areas have the same form and shape." With this bold statement and using dimensional analysis, he derived a scaling of urban area with the power $2/3$ of the population, and found a very good agreement with data from the (remarkabky precise for the time) 1960 and 1965 Swedish censuses. Even though the city definition seems not to correspond to ours, the scaling is the same since the discs defined by a constant value of our rescaled radius $r''$ have a population which is proportional to the total LUZ population and an area proportional to the power $2/3$ of the total population.

\begin{figure}[h!]
\begin{center}
\begin{tabular}{cc}
\includegraphics[height=7cm]{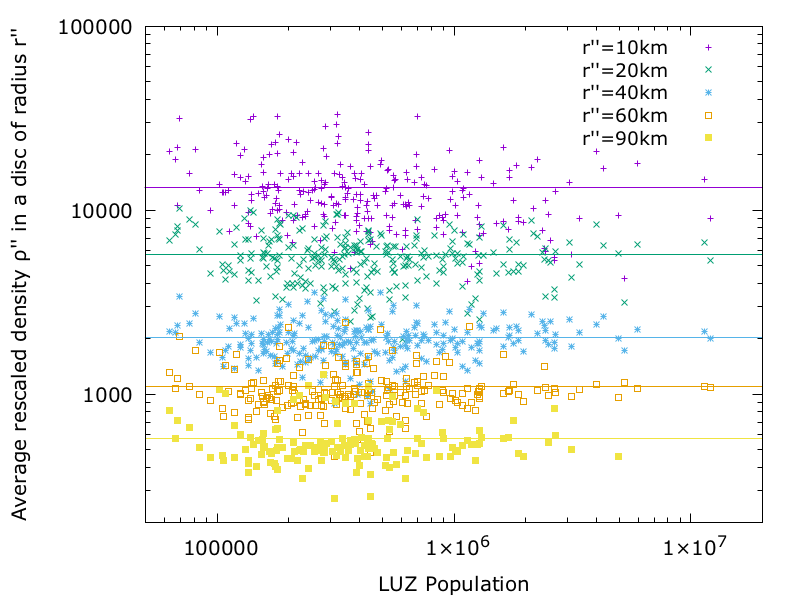} &
\includegraphics[height=7cm]{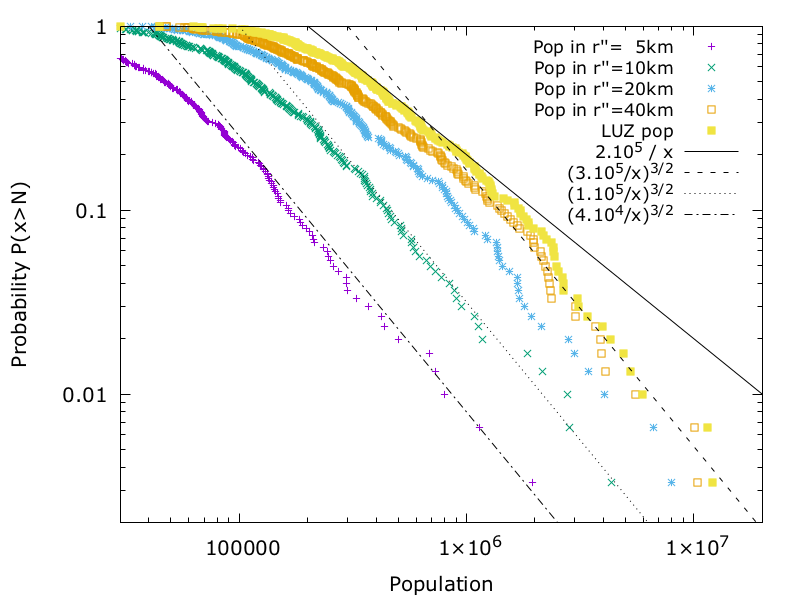}
\end{tabular}
\caption{{\small Left panel: average rescaled density $\rho''$ in a disc of rescaled radius $r''$, for all cities of the database and different values of $r''$. The horizontal lines give the average value over all cities. Right panel: counter-cumulative distribution of city sizes in the database $P(x\geq N)$, using population counts inside discs of rescaled radius $r''$, for several values of $r''$. The lines are meant as guides to the eye.}}
\label{disc_zipf}
\end{center}
\end{figure}

This is actually another possible viewpoint on the scaling of population density, which consists in studying the evolution with population size of the average population density in the discs of different sizes defined by a fixed rescaled radius $r''$. This is done on the left panel of figure \ref{disc_zipf}, which shows that the average rescaled density $\rho''$ is roughly constant in these discs. Fluctuations between cities are high, but no drift from constant rescaled density can be observed, for any rescaled radius $r''$. This provides us with a new way of defining comparable urban areas at different scales, focusing for instance only on the core city (low values of $r''$), or rather encompassing more distant periurban or rural areas (high values of $r''$).

The right panel of figure \ref{disc_zipf} studies the distribution of cities' sizes in the database. It compares this distribution for the populations of the Larger Urban Zones used in the database, to the populations of reduced urban areas based on a fixed rescaled radius $r''$. The populations of the Larger Urban Zones do not follow Zipf's law (corresponding to an exponent of $-1$ in the histogram of the right panel of figure \ref{disc_zipf}), as observed also by \cite{bettencourt2016} for European cities of more than 500 000 inhabitants. Reduced urban areas also depart from Zipf's law in the same manner. However, these cities follow quite clearly a power-law tail distribution with an exponent close to $3/2$, irrespective of the value of the rescaled radius $r''$ at which these cities are considered, and departing from the exponent $1$ expected from Zipf's law. Actually, the populations of Larger Urban Zones themselves also follow a such power-law tail distribution with exponent 3/2. This comes as a surprise, as Larger Urban Zones are considered a good homogenous definition of urban areas across European cities, and cities of different geographical zones in the world have been shown to be quite consistent with Zipf's law \cite{nitsch2005}.

We note that although London has the largest population (soon followed by Paris) across the database when considered at the level of the LUZ, Paris is clearly the first European city when considering discs defined with small values of the rescaled radius $r''$. This could simply point to the fact that the Central Business District (CBD) of Paris is not located right next to the city hall, as opposed to London. We note also that the power-law distribution observed on the histograms of the right panel of figure \ref{disc_zipf} fails to represent the distribution of smaller cities, with a LUZ population size of 500 000 and less. One probable explanation is the fact that many cities of population between 100 000 and 500 000 are missing from the Urban Atlas, maybe because of constraints at the time of data gathering. On the contrary, no major city of size 500 000 and more is missing, to our knowledge. We note also that some cities with population clearly smaller than 100 000 are included in the database, as can be seen on figures \ref{disc_artif} and \ref{disc_zipf}.

One possible reason for this observed deviation from Zipf's law for European cities lies in the way urban areas are defined in our approach: using a fixed rescaled radius $r''$ amounts roughly to using a fixed rescaled density threshold $\rho''$ to determine the border of the urban area, so that the (non-rescaled) density threshold $\rho$ scales like the cube root of city population. This is to our knowledge a new approach in the literature. On the contrary, having a fixed density threshold (like some works do in the literature) would by comparison increase the population of large cities and decrease the population of small cities, so that the power-law exponent would be smaller -- that is, closer to Zipf's law. We note also that the exponent of this distribution of city sizes has been shown to decrease over time \cite{nitsch2005}. This could be an explanation why European cities, which still bear the trace of a long history, present a high exponent. Moreover, our use of a restricted definition of cities with fixed rescaled radiuses $r''$ of low value in the right panel of figure \ref{disc_zipf} could be a way to capture former stages of development of cities, before a large part of urban sprawl occured.

\section{Discussion}

This study provides empirical evidence of the simple geometrical scaling of cities regarding the radial evolution of their land use and population density with distance to the centre. European cities show remarkably constant profiles across city size. The urban land use scales in two dimensions, like the square root of city population, and population density, scales in three dimensions, like the cube root of city population. Both scalings are homothetic, as a surface for land use and as a volume for population density.

Our findings complement more incomplete and dispersed results that appeared decades ago in the literature \cite{tobler1969, nordbeck1971}, when data were less detailed and extensive. These former works suggest that the scaling laws we present here, based on 300 European cities, are probably also valid for other areas in the World and other time periods. We confirm for instance that the urban area scales as the power $2/3$ of city population -- the result of \cite{nordbeck1971} -- when density curves are considered, but the urban area and population are proportional when considering land use.

Our work also confirms the strong monocentricity of the European city and shows that people and activities are very inhomogenously located in urban areas. A large part of urban areas is actually (semi-)natural, as figure \ref{resc-artif} shows. This clearly calls for reconsidering the aggregate/average density approaches in urban systems analysis and for drawing explicit links between scaling laws and monocentric intra-urban models. Obviously, generic urban models, that would aim to cover cities of different sizes, should be able to account for the scaling phenomena studied here and the generic profiles of land uses and population we have found.

The results we show here are generally consistent with the theoretical model of \cite{bettencourt2013}. Rather than proposing additions to it though, or rather than developing competing models, that would cover the radial evidences we show here, we think priority should be given to further testing monocentric models of the Alonso-Muth-Mills tradition. They are well established in geographic, economic and planning research and relate internal structures of the city, similar to those we have found, to explicit households or firms behaviour. Whether these models are able to respond to the scaling effects empirically demonstrated here should be the object of explicit tests in the future. An important variable, difficult to obtain for such a large dataset, is missing from our analysis in order to further tests urban economic models: the profile of land or housing prices. Finding out empirically how these profiles scale with total population is an important next step in that perspective. Urban economic models suggest that the scaling would be roughly the same as for population density but validation require more data to be assembled.

There is a lot to gain from capitalising from and further integrating intra- and inter-urban modelling literature especially by bringing more space into scaling laws research and more scaling behaviour into intra-urban research. For example, we have shown that one can actually consistently define an urban area independently of its population as soon as the scaling is established. This is beneficial to the scaling laws analyses which struggle to find a proper way to include homogeneously defined cities in their observation set. We have shown here how it affects our understanding of Zipf (figure \ref{disc_zipf}) but this can be used more generally for other city-wide attributes.
%functions and systems of cities \cite{rozenblat1993, rozenblat2007}, as well as city size distributions as discussed above regarding the left panel of and Zipf's law for cities.

It is also important for comparative research at the intra-urban scale to have comparable definitions of urban areas across sizes, that can change depending on the objective. For some studies, related for instance to land take or biodiversity, land use can be taken as the relevant phenomenon, whereas population density should give a more relevant definition of cities when social aspects are studied, such as economic output, employment, crime or social interactions.

A limitation of our work has been to use as a scaling parameter the population provided from the area covered by the Larger Urban Zone. A better approach should be to loop between the data and the scaling law from which we derive the city definition. This is an interesting perspective, provided that convergence to a fixed point can be reached. A simpler approach would be to find the best fit of each individual city to the average curve. Another interesting question would then be the difference, or consistency, between the value of population size obtained from land use scaling compared to the one obtained from the scaling of population density.

\subsection*{Acknowledgements}
RL acknowledges interesting discussions with E. Altmann, F. Simini and P. Jensen.

\appendix

\section{City center location}
\label{app1}
In this section, we study the influence of the exact location of the city center on the land use and population density curves studied here. We change arbitrarily the location of the city center from the city hall to other locations in the vicinity.
\begin{figure}[h!]
\begin{center}
\begin{tabular}{cc}
\includegraphics[height=6.0cm]{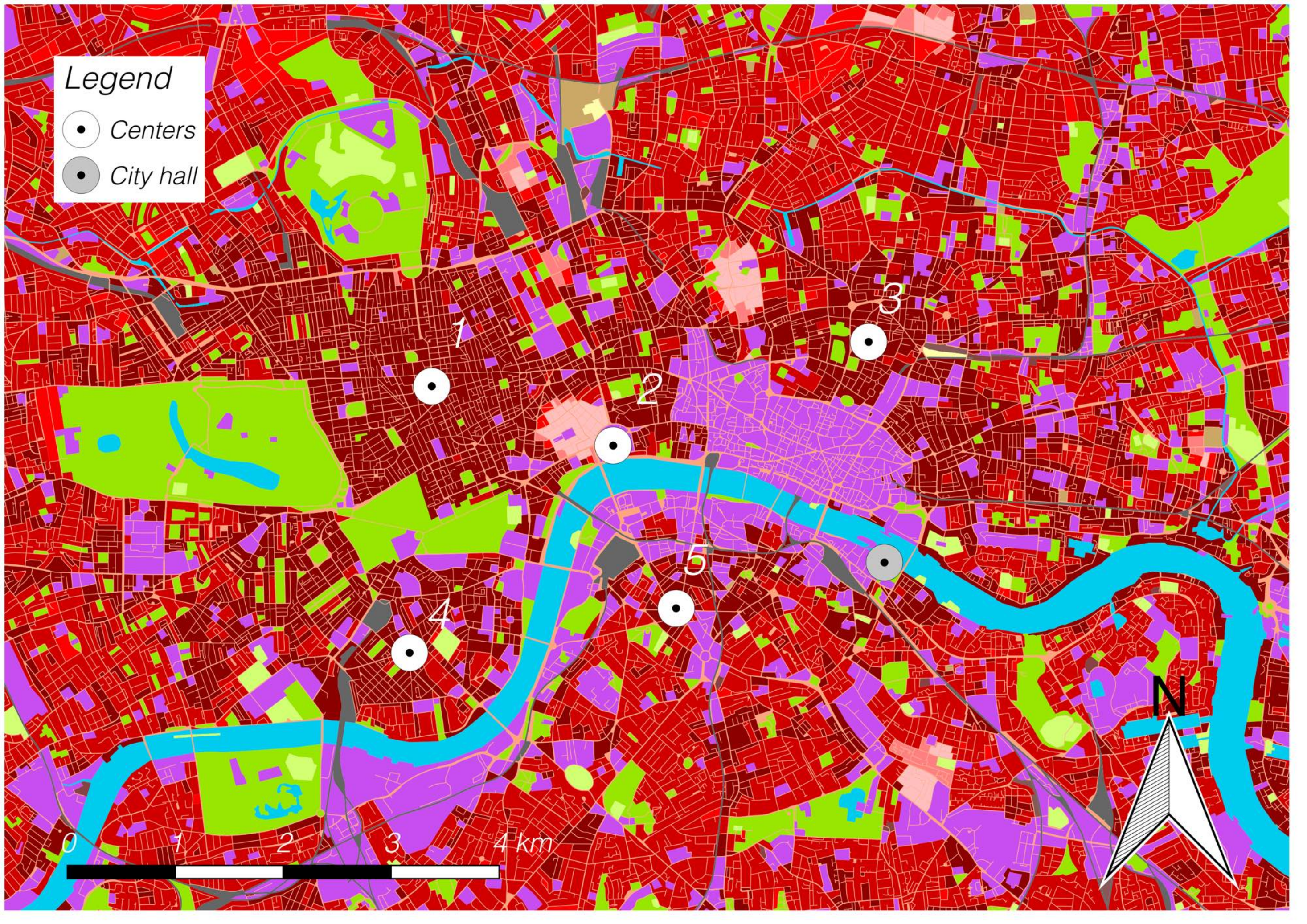} &
\includegraphics[height=6.5cm]{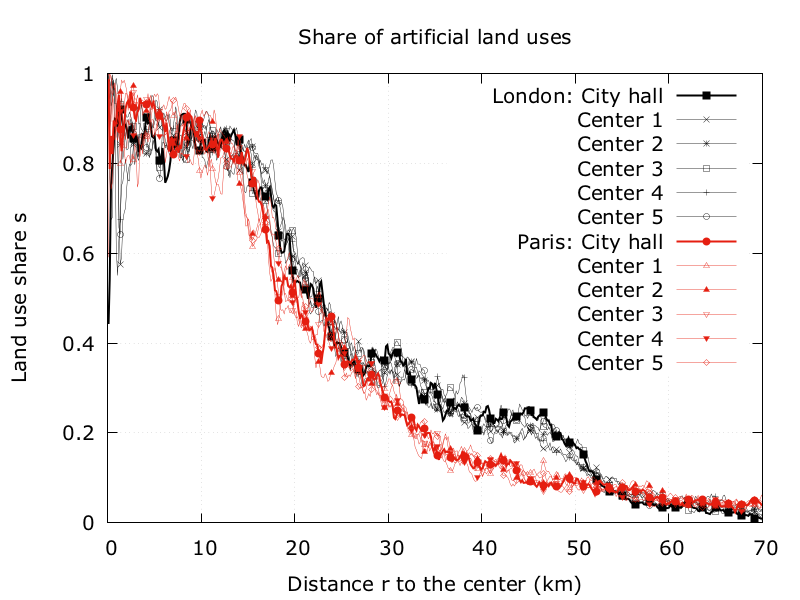} \\
\includegraphics[height=6.2cm]{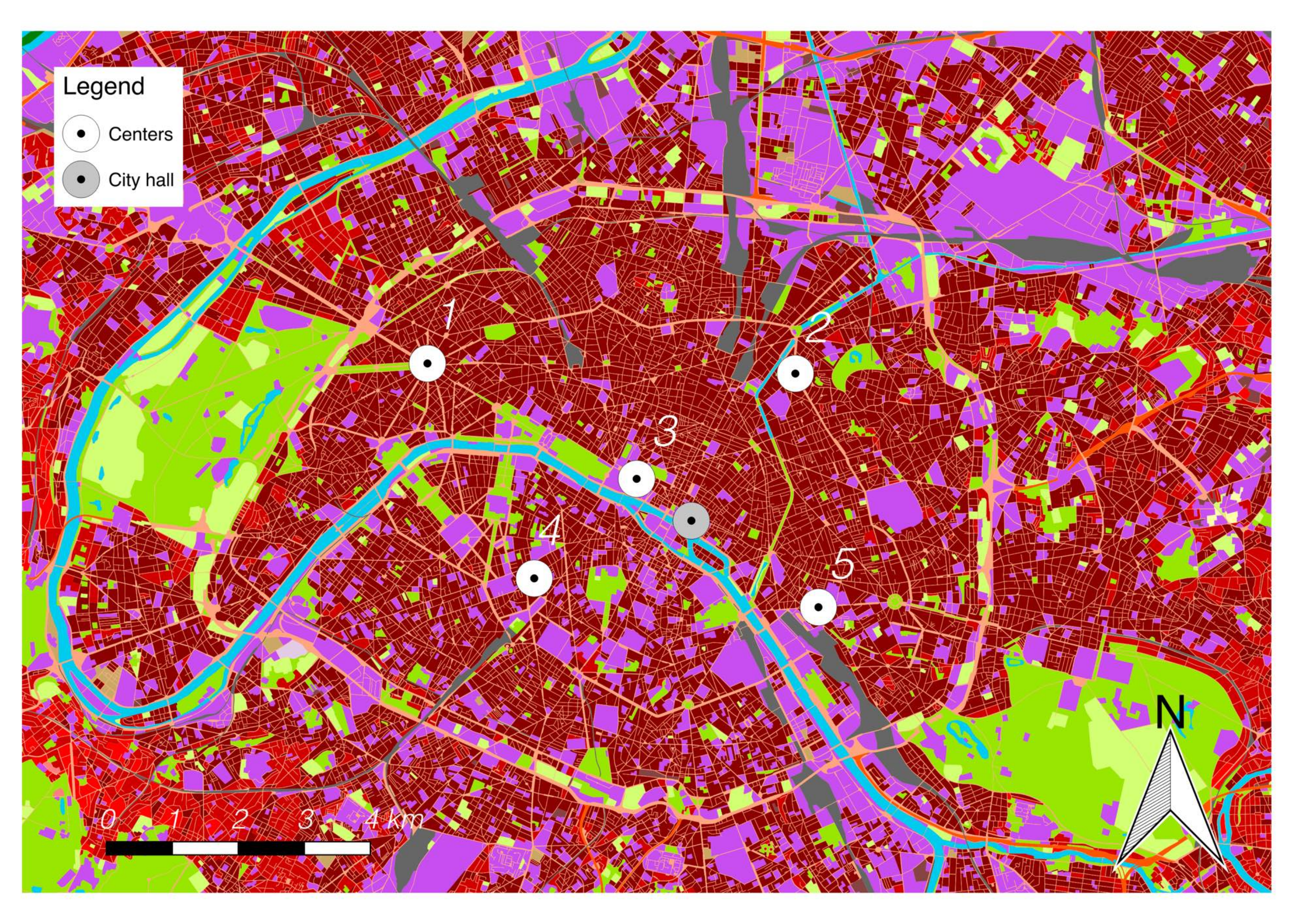} &
\includegraphics[height=6.7cm]{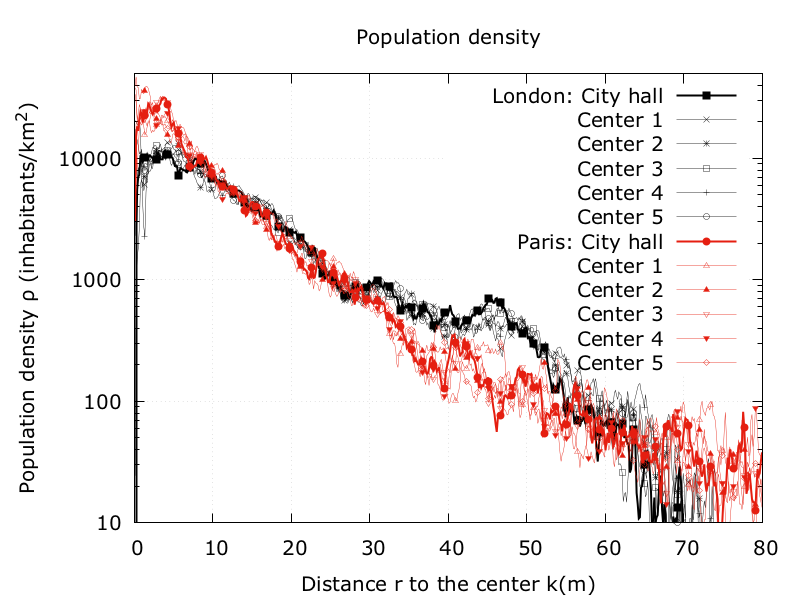} 
\end{tabular}
\caption{{\small Illustration of the influence of the precise location of the city center on the monocentric analysis. Left panel: location of the city hall and of arbitrary centers for London (top) and Paris (bottom). Right panel: influence on the artificial land use curves (top) and on the population density curves (bottom) of the location of the center.}}
\label{center}
\end{center}
\end{figure}
Figure \ref{center} shows for the two largest urban areas of the dataset, London and Paris, that displacing the center from a few kilometers has a very small influence on the results: the shape of the land use and density curves is conserved through this shift.

\bibliography{biblioLemoyCaruso.bib}
\bibliographystyle{unsrt}

\end{document}